\documentclass[10pt,aps,twocolumn,prd,noshowpacs,nofootinbib,noshowkeys,floatfix]{revtex4-1}
\usepackage[dvips]{graphics,graphicx}
\usepackage[colorlinks=true,linktocpage=true,linkcolor=blue,citecolor=blue]{hyperref}
\usepackage[usenames,dvipsnames]{color}
\usepackage{amsmath, amssymb}
\usepackage{multirow}
\usepackage{longtable}
\usepackage{color}
\usepackage[normalem]{ulem}  

\renewcommand\sout{\bgroup \color{blue} \ULdepth=-.5ex \ULset}
\newcommand{\beq}{\begin{equation}}
\newcommand{\eeq}{\end{equation}}
\newcommand{\eq}{Eq.~}

\newcommand{\linv}{\tilde{\lambda}}

\begin{document}

\preprint{}

\title{Virtual photon polarization and dilepton anisotropy in relativistic\\ nucleus-nucleus collisions}

\author{Enrico Speranza}
\affiliation{Institute for Theoretical Physics, Goethe University, D-60438 Frankfurt am Main, Germany}
\affiliation{GSI Helmholtzzentrum f\"ur Schwerionenforschung, D-64291 Darmstadt, Germany}
\author{Amaresh Jaiswal}
\affiliation{School of Physical Sciences, National Institute of Science Education and Research, HBNI, Jatni-752050, India}
\affiliation{Extreme Matter Institute EMMI, GSI Helmholtzzentrum f\"ur Schwerionenforschung, D-64291 Darmstadt, Germany}
\author{Bengt Friman}
\affiliation{GSI Helmholtzzentrum f\"ur Schwerionenforschung, D-64291 Darmstadt, Germany}

\date{\today}

\begin{abstract}

The polarization of virtual photons produced in relativistic nucleus-nucleus collisions  provides information on the conditions in the emitting medium. In a hydrodynamic framework, the resulting angular anisotropy of the dilepton final state depends on the flow as well as on the transverse momentum and invariant mass of the photon.  We illustrate these effects in dilepton production from quark-antiquark annihilation in the QGP phase and $\pi^+\pi^-$ annihilation in the hadronic phase for a static medium in global equilibrium and for a longitudinally expanding system. 

\end{abstract}



\maketitle

\section{Introduction}

In relativistic heavy-ion collisions strongly interacting matter 
at very high temperatures and densities is created~\cite 
{Rischke:2003mt, Tannenbaum:2006ch}. At such extreme conditions, 
quarks and gluons are deconfined and form a new phase of quantum 
chromodynamics (QCD), the quark-gluon plasma (QGP). Therefore 
relativistic nucleus-nucleus collisions provide a unique 
opportunity to study and characterize thermodynamic phases of QCD matter. 

Since, experimentally, the matter produced in nuclear collisions can be
probed only by observing and analyzing the spectra of 
emitted particles, it is important to understand their production 
mechanisms and interactions. Hadronic observables interact strongly 
and do not probe the entire space-time volume of the collision 
because they are emitted only from the surface and in the final state of the medium. On the 
other hand, because their mean-free paths in nuclear matter are 
much larger than nuclear sizes, electromagnetic probes such as photons and dileptons are 
emitted throughout all stages of the collision and escape from the 
system essentially without final-state interactions. Therefore 
electromagnetic radiation carries direct information on the 
space-time evolution of the matter created in such collisions 
\cite{Hoyer:1986pp, Bratkovskaya:1995kh, Shuryak:2012nf}. 

Recently, it was proposed that the polarization of real and virtual photons can be used to study the momentum anisotropy of the distributions of quarks and gluons \cite{Ipp:2007ng,Baym:2014qfa,Baym:2017qxy}.
In a first measurement of the dilepton angular anisotropy, the NA60 Collaboration found that the anisotropy coefficients in $158\,A$GeV In-In collisions are consistent with zero~\cite{Arnaldi:2008gp}, while the HADES Collaboration~\cite{Agakishiev:2011vf} finds a substantial transverse polarization in Ar-KCl at $1.76\,A$GeV. 

In general, multiply differential cross sections for dilepton 
emission provide information needed to disentangle the 
various production channels. The polarization state of the virtual photon is reflected in anisotropies of the angular 
distribution of the lepton pair. Thus, different photon production mechanisms give rise to characteristic shapes for the dilepton angular distribution \cite{Gottfried:1964nx,Falciano:1986wk,Abt:2009nu,Speranza:2016tcg}. 

The angular distribution of the leptons originating from the decay of a virtual photon, expressed in the photon rest frame, is of the form \cite 
{Gottfried:1964nx, Falciano:1986wk,Faccioli:2010kd, Faccioli:2011pn}:
\begin{align}\label{general_ang_distr0}
\frac{d\Gamma}{d^4q d\Omega_\ell}={}&\, \mathcal{N} \Big(1+\lambda_\theta\cos^2\theta_\ell \nonumber \\
&+\lambda_\phi \sin^2\theta_\ell\cos2\phi_\ell+\lambda_{\theta\phi}\sin2\theta_\ell\cos\phi_\ell \nonumber\\
&+\lambda^{\bot}_\phi \sin^2\theta_\ell\sin2\phi_\ell+\lambda^{\bot}_{\theta\phi}\sin2\theta_\ell\sin\phi_\ell \Big) ,
\end{align}
where $\Gamma\equiv\frac{d N}{d^4 x}$ is the dilepton production rate per unit volume, $q^\mu$ the virtual photon momentum while  $\theta_\ell$ and $\phi_\ell$ are the polar and 
azimuthal angles of, e.g., the negative lepton in the rest frame of the photon and $d\Omega_\ell=d\cos \theta_\ell\, d\phi_\ell$. The normalization $\mathcal{N}$ is independent of the lepton angles. 
The coefficients $\lambda_\theta$, $\lambda_{\phi}$, 
$\lambda_{\theta\phi}$, $\lambda^{\bot}_\phi$ and 
$\lambda^{\bot}_{\theta\phi}$ are the 
anisotropy coefficients, $\lambda^{\bot}_\phi$ and 
$\lambda^{\bot}_{\theta\phi}$ being non-zero only for processes that are not symmetric with respect to reflections in the production plane. 

\begin{figure}[t]
	\begin{center}
		\includegraphics[scale=1.5]{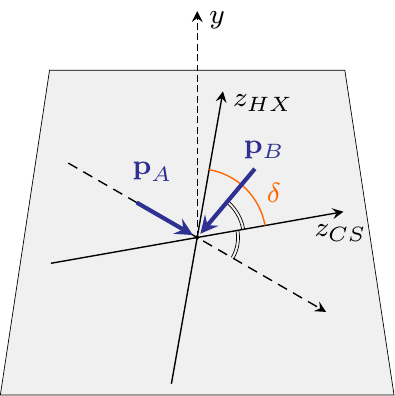}
	\end{center}
	\vspace{-0.5cm}
	\caption{(Color online) Illustration of the two reference frames employed in this paper. The production plane, indicated in gray, is spanned by the three momenta of the initial ions in the rest frame of the virtual photon, ${\bf p}_A$ and ${\bf p}_B$, and  is orthogonal to the $y$ axis. The axes $z_{HX}$ and $z_{CS}$ define the helicity and Collins-Soper frames, respectively. The two frames are connected by a rotation  through the angle $\delta$ about the $y$ axis~\cite{Faccioli:2010kd}.}
	\label{frames}
\end{figure}

The anisotropy coefficients depend on the choice of the quantization axis. In this work we employ two reference frames (see Fig.~\ref{frames}). In the helicity frame, the quantization axis is along the photon momentum, while in the Collins-Soper frame the quantization axis is the bisector of the angle formed by the beam and target momenta in the photon rest frame  \cite{Collins:1977iv,Faccioli:2010kd}. We compute dilepton emission from quark-antiquark annihilation in the QGP phase (the Drell-Yan process) and $\pi^+\pi^-$ annihilation in the hadronic phase.

The invariant mass spectrum and $q_T$ dependence of low-mass  dileptons ($M< 1$ GeV) produced in heavy-ion collisions are consistent with an equilibrated, collectively expanding source \cite{Arnaldi:2008fw,Arnaldi:2007ru}. Moreover, the lack of dilepton anisotropy found in Ref.~\cite{Arnaldi:2008gp} has been interpreted as evidence for a thermalized medium. However, as noted in Ref. \cite{Hoyer:1986pp}, also a fully thermalized medium in general emits polarized photons. In this letter, we present results on the anisotropy coefficients for dileptons emitted from a thermalized static medium and explore the effect of longitudinal (Bjorken) flow on the polarization observables.
   
While in Ref.~\cite{Bratkovskaya:1995kh} the coefficient $\lambda_\theta$ has been discussed for some
relevant reactions in vacuum as well as in a thermal system, a systematic study of the dilepton anisotropy coefficients in heavy-ion collisions has so far not been performed. Here we present a general framework for studying the angular anisotropy of dileptons emanating from the decay of polarized virtual photons produced in a hot, expanding medium. In order to characterize the angular distribution of the dilepton final state, we introduce the anisotropy coefficients and demonstrate explicitly their dependence on the flow velocity and temperature profile of the medium. We first consider the simplest case of a static uniform medium and then incorporate the non-trivial dynamics of a longitudinally expanding system.

QCD matter formed in high-energy heavy-ion collisions exhibits a striking collective behavior and its space-time evolution can be described quite accurately with relativistic hydrodynamics (for a recent review, see \cite{Florkowski:2017olj}). We use one-dimensional longitudinal Bjorken flow to account for the expansion of the medium along the beam axis. We find that in general the anisotropy coefficients in a thermalized medium are nonvanishing and depend on the flow of the medium as well as on the transverse momentum and invariant mass of the virtual photon. 

Throughout the text we use natural units, with $c = \hbar = k_B =1$.


\section{Angular distribution}

Consider the annihilation process $X_1 X_2\to\gamma^*\to \ell^+ \ell^-$, where $X_1$ and $X_2$ denote a particle and its antiparticle, while $\gamma^*$ is the intermediate virtual photon, which decays into a lepton-antilepton pair. The rate per unit volume for this process can be written as~\cite{McLerran:1984ay,Gale:1990pn}
\begin{align}\label{generalrv}
\frac{d\Gamma}{d^4q} ={}& \frac{e^4}{q^4} \int \frac{d^3 l_+}{(2\pi)^3 2E_+}\frac{d^3 l_-}{(2\pi)^3 2E_-}\,W^{\mu\nu} L_{\mu\nu}\\
&\times\,\delta^{(4)}(q-l_+ - l_-) ,\nonumber
\end{align}
where $E_{\pm}=\sqrt{|{\bf l}_{\pm}|^2+m_\ell^2}$, $m_\ell$ is the lepton mass, $e$ its charge and $L_{\mu\nu}$ the 
lepton tensor
\begin{equation}\label{eqlep}
L^{\mu\nu}=2(-q^2 g^{\mu\nu}+q^\mu q^\nu-\Delta l^\mu \Delta l^\nu).
\end{equation}
Here $l_+^\mu$ and $l_-^\mu$ are the antilepton and lepton momenta, while $q^\mu\equiv l^\mu_+ + l^\mu_-$ is the 
virtual photon momentum, and $\Delta l^\mu\equiv l^\mu_+ - l^\mu_-$. 

For dilepton emission from a medium in local thermodynamic equilibrium, the tensor $W^{\mu\nu}$, which describes the annihilation of $X_1$ and $X_2$ into a virtual photon, is given by an ensemble average
\begin{equation}\label{wspecific}
W^{\mu\nu}=\langle w^{\mu\nu} \rangle .
\end{equation}
The tensor $w^{\mu\nu}$, which describes the elementary process, will be discussed below and in Sect.~\ref{sect:2C}. The ensemble average, denoted by the brackets in (\ref{wspecific}), is of the form 
\begin{align}\label{notation}
\langle A \rangle =& \int \frac{d^3 p_1}{(2\pi)^3 2E_1} \frac{d^3 p_2}{(2\pi)^3 2E_2} (2\pi)^4 \delta^{(4)}(q-p_1-p_2) \nonumber\\
&\times \frac{1}{e^{(u\cdotp p_1)/T}\pm1} \frac{1}{e^{(u\cdotp p_2)/T}\pm1} A .
\end{align}
Here $p_1^\mu$ and $p_2^\mu$ are the momenta of $X_1$ and $X_2$, respectively, 
$E_{1,2}=\sqrt{|{\bf p}_{1,2}|^2+m^2}$, $T$ is the temperature and
\begin{equation}
u^\mu=(\gamma,\gamma \bf{v}) 
\end{equation}
is the four velocity of the medium. The plus and minus signs in the distribution functions in (\ref{notation}) apply when the particles $X_1$ and $X_2$ are fermions or bosons, respectively. 

The tensor $w^{\mu\nu}$ in Eq.~(\ref{wspecific}) is given in terms of the matrix elements of the electromagnetic current operator $J_\mu$,
\begin{equation}
\label{smallw}
w_{\mu\nu}=\sum_{\text{pol}}  \mathcal{M}^{\text{prod}}_\mu \mathcal{M}^{\text{prod}*}_\nu ,
\end{equation}
where
\begin{equation}
\mathcal{M}^{\text{prod}}_\mu=\langle 0 | J_\mu | X_1 X_2 \rangle .
\end{equation}
The sum in (\ref{smallw}) is over the spin states of $X_1$ and $X_2$.

The most general structure for $W^{\mu\nu}$ is composed of 
the metric tensor $g^{\mu\nu}$ and the two four vectors that we have at our disposal, $q^\mu$ and $u^\mu$. By requiring that $W^{\mu\nu}$ is symmetric and current 
conserving, one finds
\begin{align}\label{wgeneral}
W^{\mu\nu}={}&W_1 \left( g^{\mu\nu} - \frac{q^\mu q^\nu}{q^2} \right) \nonumber \\
&+ W_2 \left( u^\mu - \frac{u\cdotp q}{q^2}q^\mu \right)\left( u^\nu - \frac{u\cdotp q}{q^2}q^\nu \right),
\end{align}
where $W_1$ and $W_2$ are Lorentz invariant functions, similar to the structure functions in deep inelastic scattering, that depend on $q^2$, $u\cdot q$ and the temperature $T$. The Lorentz scalars in (\ref{wgeneral}) are given by
\begin{equation}\label{W1W2}
W_1 = \frac{\alpha a -\beta}{2 a} , \quad
W_2 = \frac{3\beta -\alpha a}{2 a^2}  ,
\end{equation}
where $\alpha\equiv g_{\mu\nu}W^{\mu\nu}$, $\beta\equiv u_\mu u_\nu W^{\mu\nu}$ and $a=1-\frac{(u\cdotp q)^2}{q^2}$.


\subsection{Dilepton production rate}\label{sect:2A}

Starting from Eq.~(\ref{generalrv}), we can write the angular 
distribution of the lepton pair in the photon rest frame as
\begin{equation}\label{angdistr1}
\frac{d\,\Gamma}{d^4q d\Omega_\ell}= \frac{\alpha^2}{32 \pi^4} \,\frac{1}{q^4}\,\sqrt{1-4 m_\ell^2/q^2}\,\, W_{\mu\nu}L^{\mu\nu},
\end{equation}
where $\alpha=e^2/4\pi$. In the helicity frame the polar angle is measured relative to the three momentum of the photon in a reference frame, while the azimuthal angle is measured with respect to the plane formed by the beam axis and the photon momentum, the production plane.

Using Eq.~(\ref{eqlep}) and Eq.~(\ref{wgeneral}), the angular 
distribution becomes 
\begin{align}\label{angdistr}
\frac{d\,\Gamma}{d^4q d\Omega_\ell}=&\frac{\alpha^2}{16 \pi^4}\, \frac{1}{q^4}\,\sqrt{1-4 m_\ell^2/q^2}\,\Big[(-3q^2 - \Delta l^2)W_1\nonumber\\
+&[-q^2+(u\cdotp q)^2-(u\cdotp \Delta l)^2]W_2\Big] .
\end{align}
The angular dependence is expressed in polar coordinates in the helicity frame:
\begin{equation}\label{deltal_HF}
\Delta l^\mu =(0,2|{\bf l}|\sin\theta_\ell \cos\phi_\ell,2|{\bf l}|\sin\theta_\ell \sin\phi_\ell,2|{\bf l}|\cos\theta_\ell).
\end{equation}
We note that the only dependence on lepton angles in (\ref{angdistr}) is in 
the $(u\cdotp \Delta l)^2$ term. The angular distribution is parametrized in the general form (\ref{general_ang_distr0}).

As mentioned in the introduction, the anisotropy coefficients depend on the choice of the quantization axis. 
The coefficients in different coordinate systems are related by rotations~\cite{Faccioli:2010kd}.
For a fluid cell with a given velocity $u^\mu$, the 
anisotropy coefficients $\lambda_\theta$, $\lambda_\phi$, 
$\lambda_{\theta\phi}$, $\lambda^{\bot}_\phi$, 
$\lambda^{\bot}_{\theta\phi}$ and the normalization $\mathcal{N}$ 
are obtained from Eq.~(\ref{angdistr}).


\subsection{Drell-Yan and pion annihilation}\label{sect:2C}
For the Drell-Yan process (i.e., quark-antiquark 
annihilation $q\bar{q} \to e^+ e^-$), the current is given by $J^\mu_q=\sum_f e_f\,\bar{\psi}_f\gamma^\mu \psi_f$, where $\psi$ is the quark field and $e_f$ the quark charges. Thus, the explicit expression for the $w^{\mu\nu}$ tensor of \eq\eqref
{smallw} is formally the same as the lepton tensor: 
\begin{equation}
w^{\mu\nu}=2\,C_q\,(-q^2g^{\mu\nu}+q^\mu q^\nu-\Delta p^\mu \Delta p^\nu) ,
\end{equation}
where $p_1^\mu$ and $p_2^\mu$ are the quark and antiquark momenta and $\Delta p^\mu = p_1^\mu -p_2^\mu$. The sum over quark flavors and colors yields the factor $C_q=N_c \sum_f e_f^2$, where $N_c$ is the number of colors. For two flavors and three colors $C_q=5/3$. The scalar contractions 
$\alpha\equiv g_{\mu\nu}W^{\mu\nu}$ and $\beta\equiv u_\mu u_\nu 
W^{\mu\nu}$ are given by 
\begin{align}
\alpha =&~ 2\,C_q\,(-3 q^2-\Delta p^2)\langle1\rangle , \\
\beta = &~ 2\,C_q\,\left[((u\cdot q)^2 - q^2)\langle 1 \rangle -\langle (u\cdot \Delta p)^2 \rangle \right],
\end{align}
where we use the notation introduced in \eq\eqref{notation}. 

In the case of the pion annihilation process (i.e., $\pi^+ \pi^- \to e^+ e^-$), the 
photon couples to the pion convection current $J^\mu_\pi=~  (\Phi^- \partial^\mu \Phi^+ - \Phi^+ \partial^\mu \Phi^-)$, where $\Phi^+$ ($\Phi^-$) is the field of the positive (negative) pion. For the annihilation process, the current is proportional to the difference of the momenta of the two pions $\Delta p^\mu$. One thus finds 
\begin{equation}
w^{\mu\nu}=\Delta p^\mu \Delta p^\nu,
\end{equation}
which leads to
\begin{equation}
\alpha = \Delta p^2 \langle 1 \rangle, \quad \beta = \langle (u \cdot\Delta p)^2 \rangle .
\end{equation}

We note that $\Delta p^2=4 m^2-q^2$ is a constant for a given value of the photon invariant mass $M=\sqrt{q^2}$.
Here $m$ is the mass of the incident particles.

\section{Medium and flow}

In this section, we compute the anisotropy coefficients for 
dilepton emission from a medium. We consider two cases: (i) a static medium with uniform temperature and (ii) a longitudinally expanding system with Bjorken flow.


\subsection{Static uniform medium}
\label{Sec:IIIA}
In the photon rest frame, the collective velocity of the fluid is in the direction opposite to the photon momentum in the fluid rest frame. Thus, in the helicity frame, where the $z$-axis is along the photon momentum, the fluid velocity is given by 
\begin{equation}\label{umu_static}
u^\mu = \gamma_z(1,~0,~0,~v_z)=\frac{1}{M}(E_\gamma,-{\bf q}),
\end{equation}
where $\gamma_z\equiv1/\sqrt{1-v_z^2}=E_\gamma/M$, $v_z = -{|{\bf q}|}/{E_\gamma} $, while $E_\gamma=q^0$ and ${\bf q}$ are the energy and momentum of the virtual photon in the fluid rest frame, respectively.

In the photon rest frame, the 
phase-space integrals in Eq.~(\ref{notation}) are, using the azimuthal
symmetry, reduced to an integral over the angle $\vartheta$ between the momentum of one of the 
incident particles ($q$ or $\pi^-$) and the fluid velocity 
\begin{equation}\label{mean_static}
\langle A \rangle= \frac{1}{8\pi}\int_{-1}^{1} d(\cos\vartheta) \, \chi\,
\frac{1}{e^{(u\cdot p_1)/T}\pm 1} \, \frac{1}{e^{(u\cdot p_2)/T}\pm 1} A.
\end{equation}
Here $\chi\equiv\sqrt{1-4m^2/M^2}$ and
\begin{align}
u\cdot p_1 =&  \frac{E_\gamma}{2} + \frac{|{\bf q}|}{2}\,\chi\,\cos\vartheta, \\
u\cdot p_2 =& \frac{E_\gamma}{2} - \frac{|{\bf q}|}{2}\,\chi\,\cos\vartheta.
\end{align}For a static system, the multiplicity is given by an integral of the rate $d\,\Gamma$
over space-time. This yields an overall factor $V t_e$ (volume $\times$ emission time), which cancels in the anisotropy coefficients. 

Owing to the azimuthal symmetry of the static uniform system in the helicity frame~\footnote{We stress that, for a static system, the azimuthal symmetry is a general property, independent of the elementary emission process.}, the only non-vanishing anisotropy coefficient is in this case $\lambda_\theta$, corresponding to a tensor polarized virtual photon~\cite{Leader}.  Moreover, in the limit  ${\bf q}\to 0$, the distribution functions of the initial state in Eq.~\eqref{mean_static} are spherically symmetric, leading to zero polarization and vanishing anisotropy coefficients. 

We note that in the Boltzmann limit, $u\cdot p_1,\,u\cdot p_2\gg T$, the angular dependence of the distribution functions in \eqref{mean_static} cancels, implying that $W_2$ vanishes and that the virtual photon is unpolarized~\cite{McLerran:1984ay}. Thus, the anisotropy of dileptons emitted from thermal systems in annihilation processes is a consequence of quantum statistics. It follows that the anisotropy coefficients approach zero for large photon momenta, or large invariant masses, where the distribution functions of the initial particles are well approximated by the Boltzmann distribution. 

Furthermore, for particles of non-zero mass $m$ in the initial state, the anisotropy coefficients vanish also in the limit $M\to 2 m$, since in this limit, the distribution functions in (\ref{mean_static}) are independent of the orientation of the momenta~\footnote{Since all averages defined by \eqref{mean_static} vanish for $\chi\to 0$, this is the case also for the cross section \eqref{general_ang_distr0}. However, the contributions to the dilepton anisotropy vanish with a higher power of $\chi$, implying that $\lambda_\theta\propto \chi^2$ for $\chi\to 0$.} ${\bf p}_1$ and ${\bf p}_2$.


\subsection{Longitudinal flow}

In the case of a purely-longitudinal 
boost-invariant expansion of a transversely homogeneous system first considered by Bjorken~\cite{Bjorken:1982qr}, all scalar functions of space and time depend only on the longitudinal proper 
time $\tau\equiv\sqrt{t^2-z^2}$. The fluid four-velocity is, in the center-of-momentum (c.m.) frame, given by
\begin{equation}\label{umu_bjorken}
u^\mu = \left( \cosh\eta, ~ 0, ~ 0, ~ \sinh\eta \right),
\end{equation}
where $\eta\equiv\tanh^{-1}(z/t)$ is the space-time rapidity. In the absence of viscosity, the temperature evolution of the system is given by the Bjorken scaling solution,
\begin{equation}\label{bjorken_exp}
T \propto \tau^{-1/3}.
\end{equation} 
Here a conformal equation-of-state is assumed, with $\epsilon=3P\propto 
T^4$, where $\epsilon$ is the energy density and $P$ the pressure of the system.

The photon momentum can, in the c.m.\ frame, be written as
\begin{align}\label{qmu_bjorken}
q^\mu = &( M_T \cosh y, ~ q_T\cos\phi_\gamma, ~ q_T\sin\phi_\gamma,\nonumber\\ 
&\,\,\, M_T \sinh y),
\end{align}
where $y$ is the longitudinal rapidity of the photon, $q_T$ its transverse momentum and $M_T\equiv\sqrt{M^2+q_T^2}$ its transverse mass. In the rest frame of a cell, moving with fluid rapidity $\eta$, the same photon has momentum 
\begin{align}\label{qmu_bjorken-flow}
(q')^\mu = &( M_T \cosh (y-\eta), \,q_T\cos\phi_\gamma,\, q_T\sin\phi_\gamma,\nonumber\\
 & \,\,\,M_T \sinh (y -\eta) ) .
\end{align}

For a given fluid cell, the $z$-axis of the corresponding helicity frame is aligned with the momentum of the photon in the rest frame of the cell (\ref{qmu_bjorken-flow}). Consequently, the helicity frames for photons emitted from cells with different fluid velocities do not coincide.  However, in order to define the polarization of the ensemble of photons, emitted from all fluid cells, one needs to choose a unique reference frame. A natural choice is the helicity frame of a cell, which is at rest in the c.m. frame. The $z$-axis is then aligned with the spatial part of $q^\mu$, (\ref{qmu_bjorken}). In the following this frame is denoted $HX$.  It follows that the helicity frame of a cell moving with rapidity $\eta$, $HX'$, is rotated with respect to $HX$. The rotation angle equals that between the spatial parts of $q^\mu$ and $(q^\prime)^\mu$, (\ref{qmu_bjorken}) and (\ref{qmu_bjorken-flow}). For a photon emitted transversely to the beam axis in the c.m. frame, i.e., with $y=0$, this angle is given by $\xi=\arctan\left(-(M_T/q_T)\sinh \eta\right)$. 

Another tenable choice for the reference frame is defined by choosing the $z$-axis perpendicular to the beam axis, irrespective of the direction of the photon. Clearly, this frame coincides with $HX$ for $y=0$.

The Lorentz invariant structure functions $W_1$ and $W_2$ are, for each cell, computed as in a static, uniform medium. Consequently, in the corresponding helicity frame ($HX'$), the emitted photon is tensor polarized and the only non-zero anisotropy parameter is $\lambda_\theta$. The contribution of the fluid cell to the angular distribution of the lepton pairs in $HX$ is then obtained by evaluating (\ref{angdistr}) in that frame. Clearly, the fluid four-velocity $u^\mu$ in $HX$ depends on the fluid rapidity $\eta$ in the c.m. frame. The rotation from $HX'$ to $HX$ yields non-zero contributions to $\lambda_{\phi}$ and $\lambda_{\theta\phi}$~\cite{Faccioli:2010kd}. 

Finally, the integration of the dilepton rate over the space-time evolution of the system is in a symmetric, central collision performed using
\begin{align}
\label{spacetime}
\frac{d N}{d^2 q_T d M^2 dy}&\\ = \pi R_A^2&\int_{\tau_i}^{\tau_f} d\tau\, \tau 
\int_{-\infty}^\infty d\eta \left( \frac{1}{2} \frac{dN}{d^4x d^4q} \right) ,\nonumber
\end{align}
where $\tau_i$ and $\tau_f$ are initial and final proper time of the specific phase we are considering and $R_A=1.2 A^{1/3}$ is the nuclear radius, $A$ being the nucleon number of the incident ions. The integrand in Eq.~\eqref{spacetime} depends on the integration variables $\tau$ and $\eta$ through $T$ and $u\cdotp q$.   

At this point, several comments are in order.
i) The integration over fluid cells with different rapidities in  Eq.~(\ref{spacetime}), leads to non-vanishing $\lambda_{\phi}$ and $\lambda_{\theta\phi}$ in the $HX$ frame~\footnote{For zero photon rapidity $y$, $\lambda_{\theta\phi}$ is an odd function of fluid rapidity and hence vanishes after integrating over $\eta$ with the Bjorken solution.}. Thus, not only anisotropies in the momentum distribution of the emitting particles~\cite{Baym:2017qxy}, but also collective flow is reflected in the polarization observables.  ii) Even though the Bjorken solution is boost invariant, the resulting anisotropy coefficients, defined in the reference frame $HX$, depend on the photon rapidity $y$. This is because the quantization axis in the $HX$ frame is aligned with the photon momentum in the c.m. frame (\ref{qmu_bjorken}). Nevertheless, for the photon emission rate per unit volume $d\Gamma/d^4q$, obtained by integrating Eq.~(\ref{general_ang_distr0}) over the lepton angles $\Omega_\ell$, longitudinal boost invariance is recovered. This implies that the combination $\mathcal{N}(1+\lambda_\theta/3)$, integrated over $d^4x$, is independent of $y$. 
iii) Moreover, for the Bjorken solution, the combination $\mathcal{N}(\lambda_\theta+3\lambda_\phi)$ is invariant under longitudinal Lorentz boosts.
 
We stress that the above invariances follow from the assumptions of local thermodynamic equilibrium and longitudinal boost invariance and are independent of the production process. 
As noted above, the anisotropy coefficients, $\lambda_n$ ($n=\theta,\phi$ etc.), are frame dependent, i.e., they depend on the choice of quantization axis. However, it is possible to define frame-invariant  combinations, like~\cite{Faccioli:2010kd,Faccioli:2010ej}
\begin{equation}
\label{frame:inv}
\linv \equiv \frac{\lambda_\theta+3\lambda_\phi}{1-\lambda_\phi} .
\end{equation}
In the case of the Bjorken expansion, $\linv$ is also invariant under Lorentz boosts along the beam axis.

\begin{figure}[t]
	\begin{center}
		\includegraphics[width=\linewidth]{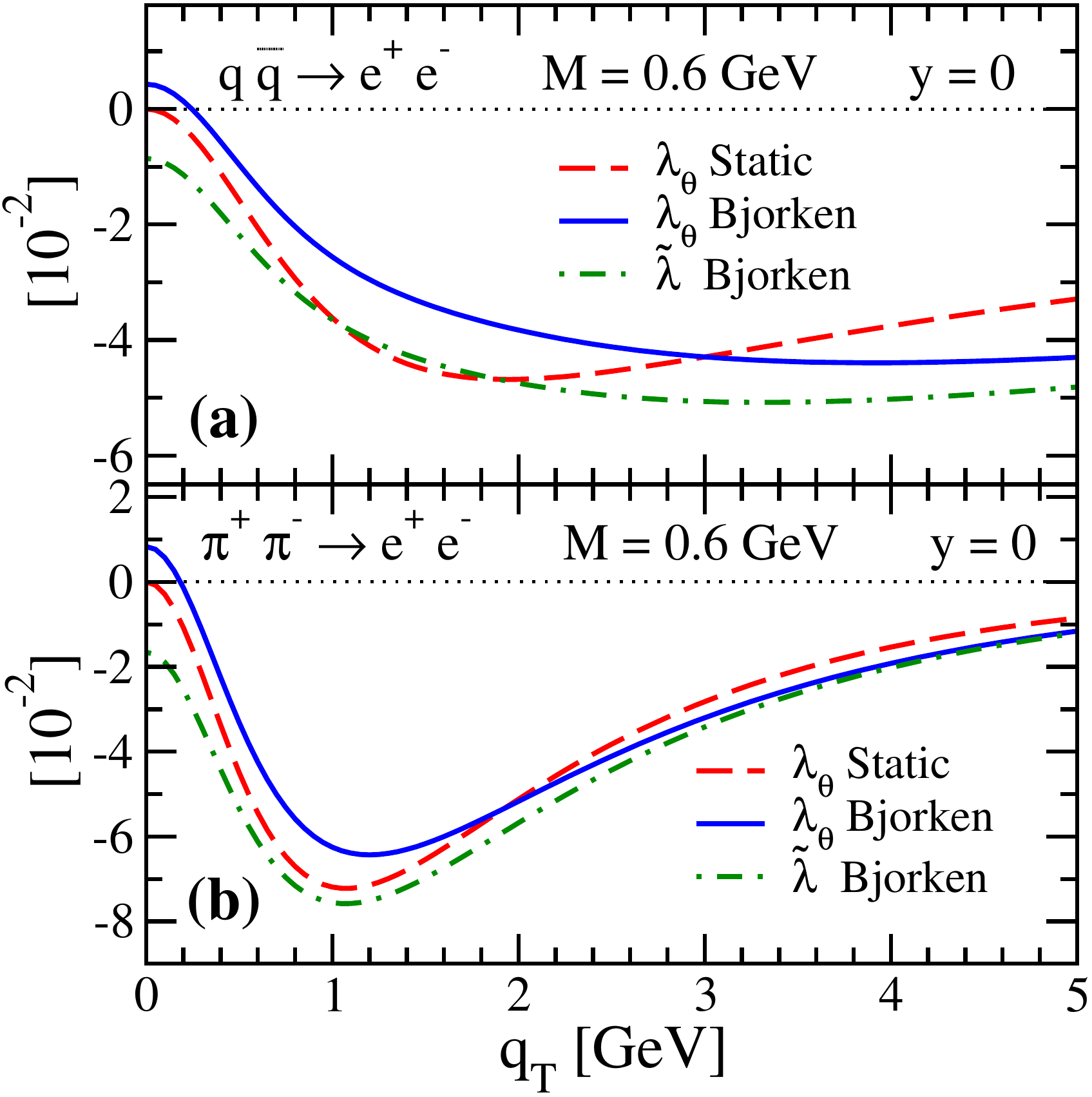}
	\end{center}
	\vspace{-0.5cm}
	\caption{(Color online) Anisotropy coefficients 
		as functions of the virtual photon transverse momentum at an
		invariant mass $M=0.6$ GeV for (a) the Drell-Yan process, and (b) pion 
		annihilation. The red dashed lines refer to $\lambda_\theta$ in the case of a static 
		uniform medium, while the blue solid lines show $\lambda_\theta$  in the helicity frame for the longitudinal 
		Bjorken expansion and the green dot-dashed lines the corresponding $\linv$.}
	\label{qT_dep}
\end{figure}


\section{Numerical results}

In this section, we present numerical results for the anisotropy 
coefficients $\lambda_\theta$ and $\linv$ for the Drell-Yan and pion-annihilation processes~\footnote{In the pion-annihilation process we use $m_\pi=139.6$ MeV, while in the Drell-Yan process we set the quark masses to zero.}. We compute these coefficients for a static uniform medium 
and for a medium with longitudinal Bjorken expansion. In the Bjorken 
case, we evolve the system from the initial temperature, $T_i=500$~MeV (at initial proper time $\tau_i = 0.4$~fm) to the final temperature $T_f=160$ MeV for the Drell-Yan process and 
from $T_i=160$ MeV ($\tau_i = 12.2$~fm) to $T_f=120$ MeV for the pion-annihilation process. In the case of a static uniform medium,
we use, in order to facilitate the comparison between the scenarios, the relevant average temperature for each process,
\begin{equation}\label{Tav_static}
T_{\rm av} = \frac{3}{2}\, T_i\, T_f \left( \frac{T_i + T_f}{T_i^2 + T_i\,T_f + T_f^2} \right),
\end{equation}
obtained using the Bjorken evolution, Eq.~(\ref{bjorken_exp}).

\begin{figure}[t]\label{fig:qtint}
	\begin{center}
		\includegraphics[width=\linewidth]{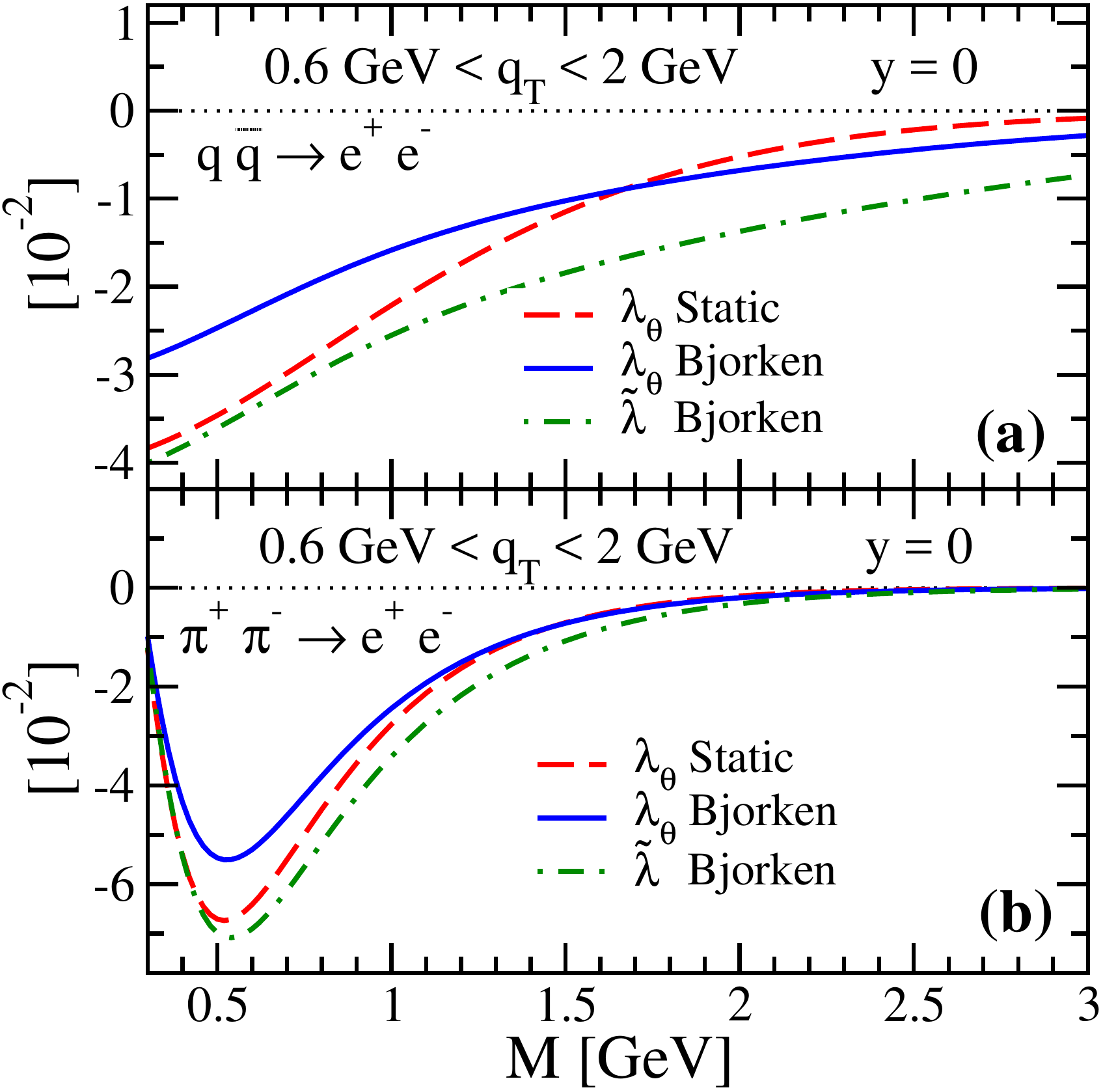}
	\end{center}
	\vspace{-0.5cm}
	\caption{(Color online) Anisotropy coefficients integrated over transverse momentum in the range between $0.6$ GeV and $2$ GeV as functions of the invariant mass $M$ for (a) the Drell-Yan and (b) the pion-annihilation processes. The red dashed lines refer to $\lambda_\theta$ in the case of a static uniform medium, the blue solid lines show $\lambda_\theta$  in the helicity frame for the longitudinal 
		Bjorken expansion, and the green dot-dashed lines the corresponding  $\linv$.}
	\label{M_dep}
\end{figure}

In Fig.~\ref{qT_dep}, we show the anisotropy coefficients 
$\lambda_\theta$ and $\linv$ against the photon transverse momentum at an invariant 
mass $M=0.6$ GeV. The coefficients are shown for the Drell-Yan and pion 
annihilation processes for two velocity profiles describing a static 
and uniform medium and a longitudinal Bjorken expansion. We observe 
that, in the static case, the anisotropy coefficient tends to zero 
for small values of the photon transverse momentum and vanishes 
at $q_T=0$ GeV, for both processes, as expected. 
 
For large values of $q_T$ the anisotropy coefficients 
again approach zero, because the momentum distribution functions 
are well approximated by the Boltzmann distribution, leading to unpolarized photons. 
We note that the anisotropy coefficients for pion annihilation 
tend to zero faster than for the Drell-Yan process, owing to the 
non-zero pion mass and the lower average temperature in the hadronic phase. 

In the case of the Bjorken expansion, 
(Fig.~\ref{qT_dep}), the anisotropy coefficients do not 
vanish in the limit $q_T\to 0$. This is a consequence of the fact  
that a photon with vanishing momentum in the c.m. frame, has a non-zero momentum in the local fluid rest frame, if emitted from a fluid element with flow. As in the static case, we observe that for large momenta the anisotropy coefficients approach zero, because the momentum distribution functions approach the Boltzmann limit. 

In Fig.~\ref {M_dep} the anisotropy coefficients, integrated over $q_T$ between $0.6$ and $2$ GeV, are shown as functions of the photon invariant mass $M$. Here the Boltzmann limit, with vanishing anisotropy, is approached for large $M$. Moreover, the pion-annihilation anisotropy coefficients vanish also in the limit $M\to 2 m_\pi$, as discussed in section \ref{Sec:IIIA} for a static medium.

Interestingly, the two processes considered yield rather similar anisotropy patterns, although the photon polarizations in the corresponding elementary reactions are distinctly different. In the Drell-Yan process, the photons are purely transverse ($\lambda_\theta=1$), while in the pion-annihilation process they are purely longitudinal ($\lambda_\theta=-1$) in a frame where the $z$-axis is along the ``beam'' axis, defined by the momenta of the incident particles in the c.m. frame. Now, when the incident particles are drawn from a Bose-Einstein distribution, momenta along the $z$ axis of the helicity frame are preferred, while for a Fermi-Dirac distribution function, there is a slight preference for momenta in the plane orthogonal to that axis. These effects conspire to yield the resulting negative values for $\lambda_\theta$ in the helicity frame, shown in Figs.~\ref{qT_dep} and \ref{M_dep}.  

In order to make a qualitative comparison with the data of NA60 \cite{Arnaldi:2008gp}, we perform an integration over the invariant mass, the photon transverse momentum and the rapidity in the intervals $0.4~\text{GeV}<M<0.9~\text{GeV}$, $0.6~\text{GeV}<q_T<2~\text{GeV}$ and $0.3<y<1.3$, respectively. To be precise, the resulting $\lambda_n$ ($n=\theta,\phi$ etc.) is the ratio of $\int\mathcal{N}\lambda_n$ and $\int\mathcal{N}$, where the integral signs indicate the integration over photon kinematics (cf. Eq.~\ref{general_ang_distr0}). In this context, we note that the relations between anisotropy coefficients in different frames~\cite{Faccioli:2010kd} are at best approximate for the ``integrated'' quantities, since the rotation angle $\delta$ (cf. Fig.~\ref{frames}) depends on the photon kinematics, i.e., on the integration variables. 

We use $T_i=250$ and $T_f=160$~MeV for the Drell-Yan process and $T_i=160$ and $T_f=120$~MeV for pion annihilation. In the helicity frame, we find  $\lambda_\theta^{HX}\simeq -0.008$ and $\lambda_\phi^{HX}\simeq -0.009$ for the Drell-Yan and $\lambda_\theta^{HX}\simeq -0.014$ and $\lambda_\phi^{HX}\simeq -0.016$ for the pion-annihilation processes. Correspondingly, in the Collins-Soper frame, the Drell-Yan process yields $\lambda_\theta^{CS}\simeq 0.002$ and $\lambda_\phi^{CS}\simeq -0.012$ and the pion annihilation process  $\lambda_\theta^{CS}\simeq 0.007$ and $\lambda_\phi^{CS}\simeq -0.023$, respectively. 

We have also computed $\tilde{\lambda}$, which is frame invariant, also when integrated over the photon kinematics. For the Drell-Yan process, we find $\tilde{\lambda}\simeq-0.034$, while for pion annihilation we obtain $\tilde{\lambda}\simeq-0.061$. We note that the integrated $\tilde{\lambda}$ in all cases differs considerably from the corresponding $\lambda_\theta$, and that the integrated $\lambda_{\phi}$ is similar in magnitude or even larger than $\lambda_\theta$. This implies that, for Bjorken flow, the $y$-dependence of the anisotropy coefficients is fairly strong both in the $HX$ and $CS$ frames and that both $\lambda_\theta$ and $\lambda_{\phi}$ are important for characterizing the polarization of virtual photons emitted from a longitudinally expanding system.


\section{Summary and outlook}

In this work we studied the polarization of virtual photons in heavy-ion collisions. In particular, we presented a general framework for studying photon polarization and the associated angular anisotropies of dileptons produced at high collision energies. We showed, using two examples, how the velocity and temperature profiles describing the evolution of the medium are reflected in the anisotropy coefficients. 

Two basic processes were considered: quark-antiquark annihilation in the QGP, and pion annihilation in the hadronic phase. We specifically studied the dilepton anisotropies for emission from a static uniform system in global thermodynamic equilibrium and a longitudinally expanding hydrodynamic system. Our results show that virtual photons originating from a static medium in thermal equilibrium are in general tensor polarized in the helicity frame. For annihilation processes, this effect is due to quantum statistics, since the polarization vanishes in the Boltzmann limit for the emitting particles. 

For both processes we found
a negative anisotropy coefficient $\lambda_\theta$, with a maximum magnitude of about $5 \%$. Integrated over $M$, $q_T$ and $y$, the coefficient $\lambda_\theta$ is on the order of $1 \%$ or smaller. The latter is compatible with the finding of the NA60 Collaboration~\cite{Arnaldi:2008gp} that the anisotropy coefficients are small, and within the experimental error, compatible with zero. Nevertheless, our results indicate that future experiments with higher statistics, could provide an unambiguous signal of virtual photon polarization effects in heavy-ion collisions, using, e.g., the frame-invariant combination  $\tilde{\lambda}$. 

In this context, we note that the large transverse polarization obtained by the HADES Collaboration~\cite{Agakishiev:2011vf} in Ar-KCl at $1.76\, A$GeV is not consistent with the annihilation processes in local thermal equilibrium considered here. The observed anisotropy may be due to non-equilibrium effects or dominated by another process, like, e.g., the $\Delta$ Dalitz decay. 

The consequences of transverse flow and viscous effects on photon polarization are under study and will be presented in a future publication. More generally, the framework presented in this paper can be easily implemented in a realistic hydrodynamic simulation of relativistic heavy-ion 
collisions in order to explore the consequences of non-trivial medium properties on the dilepton anisotropy. 
Specifically, the role of magnetic fields, fluid vorticity and non-equilibrium configurations are
problems of current interest, where photon polarization may provide useful complementary information. These problems require a straightforward extension of the formalism presented here, to allow for additional anisotropy axes. 

Finally, since the anisotropy coefficients depend on the elementary reaction, a systematic study of polarization observables for all important dilepton emission processes would be useful. In particular, the polarization of photons (real and virtual) emitted in the so called Compton processes $g + q\to q+\gamma$ and $g + q\to q+\gamma^\star\to q+\ell^+\ell^-$ (and the corresponding processes involving antiquarks) may provide information on the anisotropies of the momentum distributions of the emitting particles~\cite{Ipp:2007ng,Baym:2014qfa,Baym:2017qxy}. 

\vspace*{.5cm}
\begin{acknowledgments}
 We are grateful to Gordon Baym for very useful discussions on polarization in dilepton emission and Hans J. Specht for helpful conversations on the NA60 data. This work was supported in part by the DFG through the grant CRC-TR 211 and by the ExtreMe Matter Institute EMMI at the GSI Helmholtzzentrum f\"ur Schwerionenforschung, Darmstadt. The work of E.S. was supported by the Helmholtz YIG grant VH-NG-823 and by the BMBF Verbundprojekt 05P2015 - Alice at High Rate. A.J. is supported in part by the DST-INSPIRE faculty award under Grant No. DST/INSPIRE/04/2017/000038.
 
\end{acknowledgments}

\end{document}